\documentclass[11pt,a4paper]{article}
\usepackage{epsfig,graphicx,t1enc,amsmath}
\usepackage{amsfonts}
\usepackage[mathscr]{eucal}

\newcommand{\half}{\frac{1}{2}}
\newcommand{\be}{\begin{equation}}
\newcommand{\ee}{\end{equation}}
\newcommand{\ben}{\begin{equation*}}
\newcommand{\een}{\end{equation*}}

\newcommand{\beal}{\begin{align}}
\newcommand{\eal}{\end{align}}

\newcommand{\grad}{\nabla}

\newcommand{\eps}{\epsilon\,}
\newcommand{\epsd}{\bar{\epsilon}\,}

\newcommand{\idp}{\frac{1}{2}(1+\sigma_3)}
\newcommand{\idm}{\frac{1}{2}(1-\sigma_3)}
\newcommand{\idpo}{\frac{1}{2}(1+\sigma_3^1)}
\newcommand{\idpt}{\frac{1}{2}(1+\sigma_3^2)}

\newcommand{\idmo}{\frac{1}{2}(1-\sigma_3^1)}
\newcommand{\idmt}{\frac{1}{2}(1-\sigma_3^2)}

\newcommand{\igam}{I\gamma}
\newcommand{\gam}{\gamma}
\newcommand{\isig}{I\sigma}
\newcommand{\sig}{\sigma}

\newcommand{\pietac}{\big\{\pi,\eta \big\}^*}

\let\tilde=\widetilde

\begin{document}

\title{Events and cosmological spaces through twistors in the geometric (Clifford) algebra formalism}

\author{Elsa Arcaute\footnote{Electronic mail: e.arcaute@mrao.cam.ac.uk} \footnote{Current address: Dept. Mathematics, Imperial College London, South Kensington Campus, London, SW7 2AZ, UK.}, Anthony Lasenby\footnote{Electronic mail: a.n.lasenby@mrao.cam.ac.uk} and Chris Doran\footnote{Electronic mail: c.doran@mrao.cam.ac.uk}\\ \emph{Astrophysics group, Cavendish Laboratory,}\\ \emph{JJ Thomson Av., Cambridge, CB3 0HE, UK.}}

\date{}

\maketitle 

\begin{abstract}

Events in Minkowski space-time can be obtained from the intersection of two twistors with no helicity. These can be represented within the geometric (Clifford) algebra formalism, in a particular conformal space that is constructed from a quantum system of two particles. The realisation takes place in the multiparticle space-time algebra. This representation allows us to identify an event with the wave-function for a non-charged Klein-Gordon particle. A more general point can also be obtained, if the space is complexified and the twistors have non-zero helicity. In this case, such a point is no longer an event, but it can be identified with the wave-function for a charged Klein-Gordon particle. Other spaces of cosmological relevance can also be constructed using the 2-particle representation of the conformal space, as is the case for de Sitter and anti-de Sitter spaces. Furthermore, we show that Friedmann-Robertson-Walker spaces, with and without big bang, are fully described through two twistors, which are expressed within our formalism in terms of entities belonging to quantum mechanics: the Dirac matrix $\gamma_0$, a massless maximally entangled state and its complex conjugate.

\end{abstract}

\section{Introduction}

This work is a continuation of the work presented in \cite{twistors1}. There, we described the main physical and geometrical properties of 1-valence twistors within geometric algebra. We re-interpreted twistors as 4-d spinors, which allowed us to obtain their properties as the observables of a quantum system described by the spinor. As a result this formalism simplified enormously many key computations within the twistor theory. 

In this paper we look at some aspects of 2-valence twistors. 
These are crucial objects that are at the basis of taking the twistor space as the most fundamental entity, from which events in space-time are derived in the twistor theory. In detail, 2-valence twistors encode the incident point of two twistors. The properties of each of the incident twistors, determine whether or not the incident point corresponds to an event in space-time or whether it belongs to a more general space. In addition, the structure of the space is defined through the \emph{infinity twistors}, which are 2-valence twistors that correspond to the metric for the space.

Here we give a novel definition of the 2-valence twistors mentioned above, in terms of geometric algebra. 2-valence twistors are constructed as the outer product of two 1-valence twistors, and such a product is defined in the 2-particle space of \emph{multiparticle space-time algebra} \cite{2,STA&e} in our formalism.   
The incident point is represented in the conformal space. In the previous paper \cite{twistors1} we showed how this space is defined in the geometric algebra formalism, giving rise to the conformal geometric algebra. In order to relate the conformal point with the 2-valence twistor given by the outer product of two twistors, we need to express the incident point in the 2-particle version of the conformal geometric algebra. The construction of this representation for the conformal space determines a crucial step in our formalism. This is achieved through the massless projections of a relativistic entangled state \cite{V}. The result of such a structure is that it enables us to relate conformal points, 2-valence twistors, and relativistic states. This gives rise to a central result of this paper. We show that the incident point has the same structure as the relativistic wave-function for a spin-0 particle, according to the Bargmann-Wigner construction \cite{Barg-Wigner}. In particular, if the point belongs to the real space, and therefore represents an event in space-time, the correspondent spin-0 particle has no charge. And if the point does not correspond to an event in space-time, and therefore lives in a complexified space, it is related with the spin-0 wave-function of a charged particle. The close relation between the complex structure of the space and the charge of the particle, emerges here from the results. 

It is important to note, that although the wave-function in the Bargmann-Wigner formalism is defined in the 2-particle space, the multispinor represents \emph{a single particle and not a system of two particles}. This establishes a difference between other formalisms, where free massive relativistic spinning particles are described as a composite of massless interacting spinning particles, through the twistor phase space (see \cite{Bette'96}, \cite{Bette_etal'04} and \cite{Bette'04} for a review and a generalisation to a charged particle coupled to an external electromagnetic field). Other approaches for the construction of conformally invariant higher spin fields using the twistor space can be found in \cite{Bandos_etal'04,Bandos_etal'05}, however there is no immediate relation between these and our formalism.

Coming back to the geometric algebra approach, the structure of the 2-particle representation of the conformal space, leads to specific forms for the infinity twistors. In order to construct different cosmological spaces, such as Minkowski space-time, de Sitter and anti-de Sitter spaces, and a space with big bang, we only need three objects within our formalism to express the correspondent infinity twistors. These amount to the two massless projections of the relativistic singlet state, and to the vector $\gamma_0$ of the space-time algebra, represented by the Dirac matrix.

The work presented here is intended to be as far as possible self-contained, using the conventions and formalism already established in \cite{twistors1}. In section 2, we give the general background and results from the multiparticle space-time algebra, and the two important applications relevant to this work. These are specified in section 2.1, where we describe a maximally entangled state \cite{2,STA&e,book}, and in section 2.2, where we outline the conformal representation of the space-time in the 2-particle space \cite{V,proc}.
In section 3, we introduce the 2-valence twistors describing the incidence of twistors. Some insights into this were presented in \cite{proc}. In section 3.1, we look at the way an event in space-time emerges from the incidence of two twistors, within the conformal geometric algebra. 
In addition, we describe in this subsection how the 2-valence twistor has the same structure as the wave-function for a relativistic spinless particle. In section 3.2, we illustrate the more general case, which consists of two incident twistors that are not null, and therefore, do not give rise to a point in the real space. We show that this point instead belongs to the complexified space, and that its 2-valence twistor has an equivalent form to the wave-function for a spinless particle, which in this case has a charge. Finally in section 4, we look at the specific construction of spaces relevant for cosmology. In detail, we define the following objects in terms of the multiparticle space-time algebra: the infinity twistors for Minkowski, de Sitter and anti-de Sitter spaces, and the bang twistor giving rise to the initial singularity. This last section shows that such 2-valence twistors are the conformal representation of the points at infinity in the 2-particle background, which are given by the relativistic massless entangled states.

\section{Results from the multiparticle space-time algebra}

Quantum systems of many particles can be described through the \emph{multiparticle space-time algebra}. Furthermore this algebra can also be employed to encode degrees of freedom in quantum and in non-quantum systems. For example, later on we show that the conformal space is constructed making use of it. As a result, the conformal transformations can be related to quantum operations, and points in the space can be associated to a quantum state.

The algebra is built by taking a copy of the \emph{space-time algebra} defined in \cite{twistors1}, for each particle of the system, or for each degree of freedom. As a consequence, for an $n$-particle system we will obtain $n$ sets of basis vectors ${\{\gamma_\mu}^i\}$, that anticommute for distinct particle spaces, and satisfy
\be
\gamma_{\mu}^i\cdot\gamma_{\nu}^j = \delta^{ij}\eta_{\mu\nu}, \label{gam-MSTA}
\ee
where $i$, $j$ denote the particle space, $\mu$ and $\nu$ run from 0 to 3, and $\eta_{\mu\nu}=\text{diag}(+---)$ is the metric for Minkowski space-time. This is called the \emph{multiparticle space-time algebra} and is denoted by MSTA. It was first introduced in \cite{2} for non-relativistic particles, and in \cite{STA&e} for the relativistic case. 

From now on, the index 2 on a multivector, e.g. $M^2$, denotes the copy of the multivector $M$ in the particle space 2, while $M^1$ corresponds to the copy of $M$ in the particle space 1. For the square operation, we will use the following notation: $(M)^2$. If the quantity is a scalar, the index 2 will indicate the square operation, and therefore we will neglect the parenthesis. For example for the metric: $ds^2=dT^2+dV^2-dW^2-dX^2-dY^2-dZ^2$, the index 2 is indicating the square operation.

\subsection{Maximally entangled states}

The relativistic generalisation of a singlet state can be decomposed as a superposition of two relativistic massless states. These are essential objects within this framework, since they play a fundamental r\^ole in the description of the conformal space in the 2-particle background, and hence in the construction of the infinity twistors as well.

Next we define a maximally entangled state and give its relativistic generalisation.

The relative vectors described in \cite{twistors1} take the same form within the MSTA as in their single-particle case, viz
\be
\sig_k^a=\gam_k^a\gam_0^a,
\ee
where the super-index $a$ denotes the particle space, and $\gam_0^a$ represents the frame velocity vector. These vectors form a basis for non-relativistic systems. Restricting the system to two particles, the tensor product between them is
\be
\sig_k \otimes \sig_j \quad \leftrightarrow \quad \sig_k^1\sig_j^2 = \gam_k^1\gam_0^1\gam_j^2\gam_0^2 = \gam_j^2\gam_0^2\gam_k^1\gam_0^1 =\sig_j^2\sig_k^1.
\ee
The commutation property confirms that the geometric product of relative vectors living in different spaces in the MSTA, is equivalent to the tensor product between them. Note that from now on it will be assumed that the pseudoscalar belongs to the particle space of the multivector it is multiplying to its right, e.g. $I^1\gamma_k^1=I\gamma_k^1$ and $I^2\sig_3^2=\isig_3^2$.

The simplest construction of a 2-particle state in the MSTA, is given by the product of two single particle spinors, $\phi^1 \psi^2$.  However, this generates a space of 64 real dimensions, whereas we only expect 32 (for the 16-dimensional complex space).  The reason for this, is that we have failed to account for the fact that the quantum theory requires a single, global complex structure.  We must therefore ensure that post-multiplying the states by either $\isig_3^1$ or $\isig_3^2$ results in the same state\footnote{Recall that in \cite{twistors1} we stated that the action of the bivector $\isig_3$ to the right of the spinor, was equivalent to multiplication by the unit imaginary.}.  That is
\be
\psi \isig_3^1 =\psi \isig_3^2.
\ee
A simple rearrangement now yields
\be
\psi=-\psi \isig_3^1 \isig_3^2
=\psi \half\bigl(1-\isig_3^1 \isig_3^2 \bigr).
\ee
If we now define
\be
E=\half \bigl( 1-\isig_3^1\isig_3^2 \bigr),
\ee
we see that all states must satisfy 
\be
\psi=\psi E.
\ee
The 2-particle \textit{correlator} $E$ is an idempotent, $(E)^2=E$, that removes precisely half the degrees of freedom from a general product state, as required.  The complex structure is now provided by right multiplication by the multivector $J$, where
\be
J = E\isig_3^1 = E\isig_3^2 =
\half \left(\isig_3^1+\isig_3^2\right). \label{J}
\ee
It follows that $(J)^2=-E.$

A simple example of a non-relativistic singlet state 
\be
|\chi\rangle=\frac{1}{\sqrt{2}} 
\left \{ \begin{pmatrix} 1 \\ 0 \end{pmatrix}
\otimes \begin{pmatrix} 0 \\ 1 \end{pmatrix}
- \begin{pmatrix} 0 \\ 1 \end{pmatrix}
\otimes \begin{pmatrix} 1 \\ 0 \end{pmatrix} \right \},
\ee
can now be provided in terms of the MSTA
\be
\chi=\frac{1}{\sqrt{2}} \bigl( \isig_2^1 - \isig_2^2 \bigr) E .
\label{singlet_Pauli}
\ee
This state has the following property
\be
\isig_k^1\chi = -\isig_k^2\chi, \qquad k=1,\ldots, 3, \label{sing_cond}
\ee 
which leads to rotational invariance. In detail, for any Pauli-even multivector $M$, i.e. such that it is of the form $M=(M^0)+(M^k)I\sigma_k$ (where the coefficients $(M^0)$ and $(M^k)$ are scalars), we have
\be
M^1\chi=\tilde{M}^2\chi. \label{sing_inv1} 
\ee
Rotors, $R$, for the 3-d space are of the form of $M$, and obey $R\tilde{R}=1$, therefore the singlet state behaves as follows
\be
\chi \ \mapsto \ R^1R^2\chi=R^1\tilde{R}^1\chi=\chi, \label{sing_inv2} 
\ee
which clearly shows rotational invariance. Note that a transformation in the 2-particle representation of the conformal space takes place by applying a joint transformation in both spaces, i.e. we need to multiply by the rotor in the particle space 1: $R^1$, and by the rotor in the particle space 2: $R^2$, as shown above.

The inner product is determined within the multiparticle space-time algebra, and this indicates that it has the same form in its relativistic version. This is defined by
\be
(\psi,\phi)_s=\langle E\rangle^{-1} \left(\langle\tilde{\psi}\phi\rangle-\langle\tilde{\psi}\phi J\rangle \isig_3^a\right), \label{2()_s}
\ee
where $a$ corresponds to the particle space of the object it has to its right. Throughout this work, it is the case that $a=1$ all the time. The factor of $\langle E\rangle^{-1}$ is introduced in order to ensure that the state $|\uparrow\rangle\otimes |\uparrow\rangle$, given simply by $E$ in the MSTA, has unit norm.

The relativistic state $\zeta$ is constructed by generalising $\chi$ to a Lorentz invariant state. This is ensured by the idempotent $\half(1-I^1I^2)$ as follows. The new state  is constructed by multiplying the state $\chi$ by this projector, viz
\be 
\zeta=\sqrt{2}\chi \frac{1}{2} (1-I^1I^2). \label{singlet_rel}
\ee
Note that the factor of $\sqrt{2}$ is included in order to obtain a normalised state, viz
\be
(\zeta,\zeta)_s=1.
\ee
This new state is Lorentz invariant, since it has the following properties 
\be
M^1\zeta=\tilde{M}^2\zeta, \label{M1eta}
\ee
where $M$ is any element of the even subalgebra of the space-time algebra, which means that it has elements of grade 0, 2 and 4. This relation leads to
\be
\zeta \ \mapsto \ R^1R^2\zeta=R^1\tilde{R}^1\zeta=\zeta, \label{Lorentz_inv}
\ee
where $R$ is an even element of the space-time algebra such that $R\tilde{R}=1$, and could therefore be encoding a Lorentz transformation. This verifies Lorentz invariance.

The relativistic singlet sate can be projected into two complementary massless states lying in the ideals $\idp$ and $\idm$
\be
\zeta \idp = \epsilon, \quad
\zeta \idm = \bar{\epsilon}, \quad
\Rightarrow \ \ \zeta \ = \ \epsilon+\bar{\epsilon},
\ee
where explicitly
\beal
\epsilon&=(I\sigma_2^1-I\sigma_2^2)\idpo \idpt E ,\label{eps}\\ 
\bar{\epsilon}&=(I\sigma_2^1-I\sigma_2^2)\idmo \idmt E \label{epsd}.
\end{align}
The claim that they correspond to massless states is verified by looking at their norm
\be
(\eps,\eps)_s=0, \qquad \text{and} \qquad  (\epsd,\epsd)_s=0,
\ee
and noting that the states describing massless particles are equivalent to null vectors in Minkowski space-time. 

These two objects are mapped into one another by an operation that denotes the complex conjugation operation in the 2-spinor formalism 
\be
\epsilon=\bar{\epsilon} \sigma_1^1 \sigma_1^2.
\ee
Furthermore, they are themselves relativistic singlet states, and therefore they obey eq.(\ref{Lorentz_inv}) as well. Next we show the crucial r\^ole that they play in the 2-particle representation of the conformal space.

\subsection{Conformal representation of space-time in the 2-particle space} \label{6-d MSTA}

In this section we construct the conformal representation of Minkowski space-time in the 2-particle framework. In order to state a clear connection between our formalism and the conventional approach, we first repeat the argument given in \cite{PRII} for the construction of a 6-d pseudo-Euclidean space $\mathbb{E}^6$, and then we present its description in the MSTA.

The two extra dimensions of the 6-d space are denoted by $V$ and $W$, and hence the metric of this space corresponds to
\be
ds^2=dT^2+dV^2-dW^2-dX^2-dY^2-dZ^2, \label{metricE6}
\ee
where $T$, $V$, $W$, $X$, $Y$ and $Z$ are scalar quantities.
The relation between the points of the 4-d space and those of this space is given by taking only the points on the null cone of $\mathbb{E}^6$. The equation for the null cone of the origin is
\be
T^2+V^2-W^2-X^2-Y^2-Z^2=0.
\ee
By intersecting this cone with the plane $V-W=1$ we obtain a space identical to Minkowski space, where the metric is given by
\be
ds^2=dT^2-dX^2-dY^2-dZ^2. \label{metricM}
\ee
The extra coordinates can be expressed in terms of the others as follows
\be
V=W+1=\frac{1}{2}(1-T^2+X^2+Y^2+Z^2).
\ee
A position vector in the 4-d space $r=t\gamma_0+x\gamma_1+y\gamma_2+z\gamma_3$ can be represented in $\mathbb{E}^6$ using the property that every generator of the conic, corresponding to a line in that space, intersects Minkowski space at a unique point or at points corresponding to infinity for $V=W$. The coordinates of the 4-d space are therefore (details can be found in \cite{PRII})
\be
 t=\frac{T}{V-W}, \quad x=\frac{X}{V-W}, \quad y=\frac{Y}{V-W}, \quad z=\frac{Z}{V-W} .\label{4d-6d} 
\ee 

Let us now proceed to reproduce this construction in terms of the 4-d spinors defined in \cite{twistors1} and the MSTA (see \cite{proc} and \cite{V}).

The conformal representation of a point in the space can be constructed by taking the point at the origin in the conformal space, and then translating it making use of a rotor $T_r$ (see \cite{book}). In the single particle space, the point at the origin belonging to the conformal space, is given by a null vector, that is chosen to be one of the two null vectors generators of the conformal space: $\bar{n}$ (see \cite{twistors1}). We repeat this procedure here, although now, in this multiparticle quantum background, the conformal space is constructed from the 2-particle relativistic space. Since the singlet state $\zeta$ is given in terms of two massless states $\eps$ and $\epsd$, we can choose one of these two massless states to represent the point at the origin, and we take this to be $\epsd$. Its complex conjugate $\eps$ will be identified with the point at infinity, which in terms of twistors, corresponds to the infinity twistor giving rise to the metric for Minkowski space-time.

In \cite{twistors1} we derived the conformal representation for translations on 4-d spinors. These can also be expressed in terms of a generator, which defines a map of 4-d spinors into 4-d spinors
\be
\phi \ \mapsto \ \hat{r}(\phi)\equiv -r\phi I\gamma_3 \frac{1}{2}(1+\sigma_3), \label{op_trans}
\ee
where $r$ is the position vector in 4-d, and $\phi$ is a 4-d spinor. Using this map, we see that we can write the translation rotor as $T_r=1+\hat{r}$. Furthermore, noting that $(\hat{r})^2 = 0$, the operator can be expressed as $T_r=e^{\hat{r}}$, leading to
\be
T_r(\phi)=e^{\hat{r}}\phi=\phi-r\phi I\gamma_3 \frac{1}{2}(1+\sigma_3). \label{Tr}
\ee

If we now take the point at the origin in the 2-particle system, given by $\epsd$, and translate it to $r$ using the 2-particle version of $T_r$, we get the following object
\be
\psi=T_{r^1}T_{r^2} \ \epsd=e^{\hat{r}^1}e^{\hat{r}^2} \epsd, \label{Tr1Tr2}
\ee 
which gives the conformal representation of the 4-d position vector $r$ in the 2-particle space. The new vector corresponds to a conformal null vector. This is a consequence of the construction, since conformal transformations preserve angles, and hence the inner product is conserved, which means that $\psi$ has vanishing norm
\be
(\psi,\psi)_s=(\epsd,\epsd)_s=0. 
\ee
Using now the relation between the coordinates of $\mathbb{E}^6$ and $\mathbb{M}$, the general point in 6-d is
\be
\psi_r = (V-W) \, e^{\hat{r}^1} e^{\hat{r}^2} \epsd, \label{compact_psi}
\ee
or explicitly
\be
\psi_r=(V-W) \left( -|r|^2 \epsilon - r^1 \epsd I\gamma_3^1 -  r^2 \epsd I\gamma_3^2 + \epsd
\right). \label{Psi_p} 
\ee

In the following section we derive this element from the intersection of two rays, which encode the simplest geometric representations of twistors.

\section{Incidence of twistors}

Events are objects derived non-locally from twistors. The intersection of two loci describing geometrically a twistor each, will represent a point in the complexified space-time. This will be an event in Minkowski space-time if the loci are given by real rays. Therefore, it will be an event if both twistors are null.

The representation of twistors in terms of geometric algebra was described in \cite{twistors1}. Let $Z$ and $X$ be two different twistors, that can be expressed as follows
\beal
Z=T_{-r}(\psi)&=\psi+r\psi \igam_3 \idp, \label{Z} \\
X=T_{-r}(\phi)&=\phi+r\phi \igam_3 \idp, \label{X}
\end{align}
where $\psi$ and $\phi$ are 4-d spinors given by
\beal
\psi&=\omega \frac{1}{2}(1+\sigma_3)+\pi I\sigma_2\frac{1}{2}(1-\sigma_3), \label{Dirac_twistor} \\
\phi&=\xi \frac{1}{2}(1+\sigma_3)+\eta I\sigma_2\frac{1}{2}(1-\sigma_3). 
\end{align}
This is the Weyl representation, which implies that $\omega$, $\pi$, $\xi$ and $\eta$ are Pauli spinors.

\subsection{Event in the space-time}

The 2-valence twistor encoding the incidence of two twistors is defined as the outer product of the two twistors (see \cite{PRII}). In the previous section we saw how the outer product between spinors is defined within the MSTA. The 2-valence twistor denoted by $R^{12}$ in the geometric algebra framework, is defined as follows
\be
R^{12}=(Z^1X^2-X^1Z^2)E, \label{R12}
\ee
where $E$ is the correlator operator of the 2-particle space, and the indices 1 and 2 denote the particle space.

If $X$ and $Z$ are null, $R^{12}$ represents an event if their respective rays intersect
\be
\langle \tilde{X}Z\rangle_s=0.
\ee
This orthogonality condition is the \emph{incidence relation}.

For the case of null twistors such that
\be
\omega=0, \qquad \xi=0 ,
\ee
the quantities will take a subindex $N$. These special null twistors are represented by null rays passing through the origin, which is therefore the intersecting point in this situation. The 2-valence twistor in this case is denoted by: $R^{12}_N$. This gives the position vector in the conformal space, of the event relative to the general point at which the spinor fields are taken. If the observer lies at a general point $r$, since the intersection takes place at the origin, the position vector encoded in $R^{12}_N$ should be given by $-r$. Let us now show that this 2-valence twistor actually is the conformal representation of such a point in the 2-particle space, by comparing it with $\psi_r$ defined in the previous section. 

In the conventional setting, the 2-valence twistor is given by (this is eq.(6.2.18) in p.65 of \cite{PRII})
\be
R^{\alpha\beta}=Z^{\alpha}X^{\beta}-X^{\alpha}Z^{\beta}=\pi_{D'}\eta^{D'}
\left( \begin{array}{ll} -\half\eps^{AB}r_cr^c & i{r^A}_{B'} \\ -i{r_{A'}}^B & \eps_{A'B'} \end{array} \right).
\ee
Let us express it in terms of components in order to relate the conventional and the geometric algebra approaches. This is done by looking at the correspondence between primed and unprimed indices: $Z^2=Z_{0'}$ and $Z^3=Z_{1'}$, which leads to
\beal
R^{AB}&=\pi_{D'}\eta^{D'}(-\frac{1}{2}\epsilon^{AB}r_cr^c) \label{Rabmatrix1}\\
{R^A}_{B'}&=\pi_{D'}\eta^{D'}i{r^A}_{B'} \label{Rabmatrix2}\\
{R_{A'}}^B&=\pi_{D'}\eta^{D'}(-i{r_{A'}}^B) \label{Rabmatrix3}\\
R_{A'B'}&=\pi_{D'}\eta^{D'}\epsilon_{A'B'} \label{Rabmatrix4}
\end{align}
where $A$, $B$, $A'$, and $B'$ run from 0 to 1.

These components correspond in the geometric algebra formalism, to the projections of $R^{12}_N$ into the different ideals
\beal
R_N^{12}\idpo\idpt & =  -\eps |r|^2 \ \pietac \label{R0_00} \\
R_N^{12}\idpo\idmt & =  r^1\epsd \igam^1_3 \ \pietac \label{R0_01} \\
R_N^{12}\idmo\idpt & =  r^2\epsd \igam^2_3 \ \pietac \label{R0_10} \\
R_N^{12}\idmo\idmt & =  \epsd \ \pietac \label{R0_11}
\end{align}
If we decompose $\psi_r$ given by eq.(\ref{Psi_p}) in terms of these ideals, we get
\beal
\psi_r\idpo\idpt & =  -\eps |r|^2 \ (V-W) \\
\psi_r\idpo\idmt & =  -r^1\epsd \igam^1_3 \ (V-W) \\
\psi_r\idmo\idpt & =  -r^2\epsd \igam^2_3 \ (V-W) \\
\psi_r\idmo\idmt & =  \epsd \ (V-W) 
\end{align}
Comparing $R^{12}_N$ and $\psi_r$, we see that if the following correspondence holds
\be
\big\{\pi,\eta \big\}^* \quad \leftrightarrow \quad V-W, \label{normVW}
\ee
the two are equivalent: $\psi_r$ stands for the conformal representation of $r$ in the 2-particle space and $R^{12}_N$ for that of $-r$. The correspondence  given by eq.(\ref{normVW}) is consistent, and can be further justified as follows. $V-W \neq 0$ has to be valid in order to obtain a finite point according to eq.(\ref{Psi_p}). And the condition $\big\{\pi,\eta \big\}^* \neq 0$ ensures that $\pi$ and $\eta$ are not proportional. This last condition ensures that the two null rays are not parallel and therefore intersect at a finite point. To see this, recall that $\pi$ and $\eta$ are the respective projection parts of the twistors, and the flagpole direction of the projection part of a null twistor gives the direction of its ray. Eq.(\ref{normVW}) has therefore been verified.

$R_N^{12}$ can be written in compact notation (the equivalent of eq.(\ref{compact_psi}) for $\psi_r$) as follows
\be
R_N^{12}=e^{-\hat{r}^1}e^{-\hat{r}^2}\epsd\pietac,
\ee
or explicitly as
\be
R_N^{12}=(-|r|^2\eps +r^1\epsd \igam_3^1+r^2\epsd\igam_3^2+\epsd)\pietac.
\ee
In the particular example that we considered, the rays intersected at the origin. If we want the intersection to take pace at a general point $q$, we just need to translate the origin to that point. Under this translation the 2-valence twistor becomes
\be
\begin{split}
R_q^{12}&=T_{q^1}T_{q^2}R_N^{12} \\
&=\left(-|r-q|^2\eps +(r-q)^1\epsd \igam_3^1+(r-q)^2\epsd\igam_3^2+\epsd \right)\pietac.
\end{split}
\ee
If we now set the observer at the origin: $r=0$, we recover $\psi_q$. 
Furthermore, according to equations (\ref{R12}), (\ref{Z}) and (\ref{X}), the 2-valence twistor can be expressed in terms of the 4-d spinors: $\psi$ and $\phi$, in the following way if the fields are taken at the origin
\be
R_{\text{org}}^{12}=(\psi^1\phi^2-\phi^1\psi^2)E. \label{R_org}
\ee
Note that this is equivalent to set the observer at the origin.

In terms of the intersection point $q$, which corresponds to an event in space-time, this is
\be
R_{\text{org}}^{12}=T_{q^1}T_{q^2} \ \epsd\pietac. \label{R_orgq}
\ee

The construction given above enables us to express any general 2-valence twistor $R^{12}$ in terms of its value at the origin
\be
R^{12}=e^{-\hat{r}^1}e^{-\hat{r}^2}R_{\text{org}}^{12}, \label{R_R_org}
\ee
where $R_{\text{org}}^{12}$ is given by eq.(\ref{R_org}), and $r$ denotes the position vector of the general observer.

The nature of the \emph{incidence of twistors} within the geometric algebra framework has been elucidated. If the massless particles encoded by the twistors $Z$ and $X$ are spinless, and the projection parts of these are not proportional to each other, a conformal representation in the 2-particle space of a point in Minkowski space-time can be constructed according to eq.(\ref{R_org}), (\ref{R_orgq}) and (\ref{R_R_org}). The point is an \emph{event} in space-time, because it belongs to the \emph{real} Minkowski space. This reality condition and the structure of the 2-valence twistor can be linked with a wave-function for a spinless particle that is not charged.

In \cite{proc} we reproduced the formalism of Bargmann and Wigner \cite{Barg-Wigner} and adapted it for a spin-0 particle. The argument is the following. We take a 2-particle relativistic wave-function, which is a function of the four space-time coordinates $x^\mu$
\begin{equation}
\Psi = \Psi(x^\mu) E,
\end{equation}
that satisfies the Dirac equation for each of the particle spaces
\begin{equation}
\grad^1  \Psi \igam_3^1 = m \Psi, \quad \text{and} \quad \grad^2 \Psi \igam_3^2 = m \Psi.
\end{equation}
From the most general relativistic 2-particle state, we take the antisymmetric part. This corresponds to
\be
\Psi=(\alpha+I^1 \beta)\eps+(\theta+I^1\mu)\epsd+(u^1-Iv^1) \epsd \igam_3^1 +(u^2-Iv^2) \epsd \igam_3^2, \label{Psi}
\ee
where $\alpha$, $\beta$, $\theta$ and $\mu$ are scalars, and $u$ and $v$ are vectors. If $\Psi$ satisfies the Dirac equation, we obtain: $\alpha=\theta$ and $\beta=-\mu$. Furthermore, we recover the Klein-Gordon equation for the complex scalar field defined by: $\varphi=\alpha+I\beta$, with potential: $V=u+Iv$. Note that the same wave-function can be used for the massless case.

If the field is uncharged: $\varphi=\alpha$, we see that the wave-function given by eq.(\ref{Psi}) and the 2-valence twistor $R^{12}_N$ representing an event in Minkowski space-time can be identified! The reality condition for the locus represented by $R^{12}$ is therefore equivalent to require an uncharged Klein-Gordon field in $\Psi$.

\subsection{Point in the complexified space}

A more general locus can be constructed through the 2-valence twistor if no reality condition is imposed. This corresponds to a point in the complexified space defined by $\mathbb{C}\mathbb{M}$. In this case, the twistors need not to be null nor to satisfy the incidence relation. The locus is obtained by simply demanding that the primary parts of the twistors vanish.

Within our real algebra framework, we need to define a structure that accounts for the complexity of the space. Such a structure is achieved through the pseudoscalar of the algebra for multivectors. The `complex space' is therefore obtained by replacing the position vector by its `complexified' version, viz
\be
r \ \rightarrow \ r+Is,
\ee
where $I$ is the pseudoscalar,  $r=t_r\gamma_0+x_r\gamma_1+y_r\gamma_2+z_r\gamma_3$ and $s=t_s\gamma_0+x_s\gamma_1+y_s\gamma_2+z_s\gamma_3$.
The equation for the twistor $Z$ therefore becomes
\be
Z \ \rightarrow \ Z_c=\psi+(r+Is)\psi \igam_3 \idp,
\ee
and similarly $X$ becomes $X_c$. The locus in the complexified space is encoded in the 2-valence twistor, which is given by
\be
R^{12}_c=(Z_c^1 X_c^2-X_c^1Z_c^2)E.
\ee
This can be found by requiring the primary parts of the twistors to vanish. These are given by (see \cite{twistors1})
\be
\omega_P=Z_c\idp \quad \text{and} \quad \xi_P=X_c\idp,
\ee
and they correspond to 2-spinors, which means that they are of the form: $(a^0+a^k\isig_k)\idp$, where the coefficients $a^0$ and $a^k$ are scalars. The spinors will therefore vanish if their coefficients vanish, and this leads to the following system of equations
\beal
\langle Z_c \idp \rangle =0 \\
\langle X_c \idp \rangle =0 \\
\langle Z_c \idp \rangle_{I\sigma_k} =0 \\
\langle X_c \idp \rangle_{I\sigma_k} =0
\end{align}
where $k=1,\ldots, 3$. This is a system of 8 equations with 8 variables: $t_r,\ x_r, \ y_r, \ z_r$ and $t_s, \ x_s, \ y_s, \ z_s$, and the solutions are denoted by $r_{\text{sol}}$ and $s_{\text{sol}}$. The complex point defined by $k=r_{\text{sol}}+Is_{\text{sol}}$ encodes the position vector of the locus from an observer at the origin.
Let us now verify that the 2-valence twistor at the origin is the conformal representation of such a point.

The projections of $R^{12}_{\text{org}}$ into the different ideals are
\beal
R_{\text{org}}^{12}\idpo\idpt&=-k\bar{k} \ \eps \ \pietac \label{R_org_c1} \\
R_{\text{org}}^{12}\idpo\idmt&=-k^1 \ \epsd \  I\gamma_3^1 \  \pietac\\
R_{\text{org}}^{12}\idmo\idpt&=-k^2 \ \epsd \ I\gamma_3^2 \  \pietac\\
R_{\text{org}}^{12}\idmo\idmt&=\epsd \ \pietac
\end{align}
where $k=r_{\text{sol}}+Is_{\text{sol}}$ and $\bar{k}=r_{\text{sol}}-Is_{\text{sol}}$. These equations show that $R_{\text{org}}^{12}$ is actually the conformal representation of $k$ in the 2-particle space. Note that the coefficient $|r|^2$ in eq.(\ref{R0_00}) was replaced by $k\bar{k}$ in eq.(\ref{R_org_c1}), since we need to take into account the complex structure of $\mathbb{C}\mathbb{M}$. The 2-valence twistor can therefore be expressed as follows
\be
R_{\text{org}}^{12}=T_{k^1}T_{k^2} \ \epsd\pietac.
\ee
If the observer lies at any point $r+Is$ in $\mathbb{C}\mathbb{M}$, the general 2-valence twistor is
\be
R_c^{12}=e^{-\widehat{(r^1+I^1s^1)}} \ e^{-\widehat{(r^2+I^2s^2)}} \  R_{\text{org}}^{12},
\ee
where the `hat' over the complex vector $r+Is$ denotes the operator of eq.(\ref{op_trans}). 
The twistor has therefore the same form in both spaces, it is just a matter of specifying the position vector in the real space $\mathbb{M}$ or in the complex space $\mathbb{C}\mathbb{M}$ in the translation rotor.

Let us re-express the twistor, making use of the following
\beal
-k\bar{k}&=a+I  b, \\
\pietac&=c+d  \isig_3,
\end{align}
where $a,b,c,d \in \mathbb{R}$, and are such that
\beal
a&=-|r_{\text{sol}}|^2+|s_{\text{sol}}|^2, \\
b&=-2 \ r_{\text{sol}}\cdot s_{\text{sol}}.
\end{align}
$R_{\text{org}}^{12}$ can as a result be expressed in the following form
\be
R_{\text{org}}^{12}=(\varsigma+I^1\nu) \eps +\bar{V}^1\epsd\igam_3^1 +\bar{V}^2\epsd\igam_3^2 +(c-I^1d)\epsd. \label{R_cc_wave}
\ee
The scalars $\varsigma$ and $\nu$ are given by
\beal
\varsigma&=ac-bd, \\
\nu&=ad+bc,
\end{align}
and the `complex vector' $V$ is defined as follows
\be
V=w+Iq, \quad \text{and} \quad \bar{V}=w-Iq,
\ee
where
\beal
w&=-c \ r_{\text{sol}}+d \ s_{\text{sol}}, \\
q&=d \ r_{\text{sol}}+c \ s_{\text{sol}}.
\end{align}
Note that we used the following properties of the massless states
\be
\eps \isig_3=I\eps, \quad \text{and} \quad \epsd \isig_3=-I\epsd,
\ee
in order to simplify eq.(\ref{R_cc_wave}). We can now identify $R_{\text{org}}^{12}$ given by eq.(\ref{R_cc_wave}) with the spin-0 wave-function given by eq.(\ref{Psi}), if the following correspondence within the scalars holds
\beal
\alpha=\varsigma, \quad \beta=\nu, \\
\theta=c, \quad \mu=-d,
\end{align}
and the following one within the vectors holds
\be
u=-c \ r_{\text{sol}}+d \ s_{\text{sol}}, \quad \text{and} \quad v=d \ r_{\text{sol}}+c \ s_{\text{sol}}.
\ee
The above generalisation to the complexified space $\mathbb{C}\mathbb{M}$ is therefore related to allowing the Klein-Gordon field to be charged. \\

So far we have focused on Minkowski space-time. In order to work in different spaces, we need to introduce a metric into the space. This can be achieved through the \emph{infinity twistors}. In the next section we see how to construct these, and their cosmological applications.

\section{Cosmological spaces}

The cosmological spaces that we will contemplate in this section, are special cases of the Friedmann-Robertson-Walker (FRW) spaces (see for example \cite{H&E}). These are of interest because all FRW spaces are isotropic, i.e. they have the property of being spherically symmetric about every point. We will therefore construct the equivalent of the infinity twistor within geometric algebra for Minkowski, de Sitter and anti-de Sitter spaces. 
Furthermore, to fully describe the cosmological structure of the space, the initial singularity when present has to be accounted for. This is the case of Minkowski spaces, where a \emph{bang twistor} is introduced. For the de Sitter and anti-de Sitter spaces, there is no initial singularity, but by introducing two other twistors, the full structure of the space is given.

\subsection{Minkowski}

The infinity twistor introduces a `metric' structure into the space. This means that it gives the necessary conditions for the 2-valence twistor to represent a point in the space, and a distance can also be determined in terms of it. For the Minkowski case, this is given by (see \cite{PRII})
\be
\mathbb{I}^{\alpha\beta} = \left( \begin{array}{cc} \epsilon^{AB} & 0 \\ 0 & 0 \end{array} \right) \label{inf_twistor}
\ee
and it defines the structure of the space in the following way. If the point $R$ is real (i.e. if $\bar{R}_{\alpha \beta}  \propto  R_{\alpha \beta}$), it represents a finite point if the following inner product is non-null
\begin{equation} 
R^{\alpha\beta}\mathbb{I}_{\alpha\beta} \ = \ 2 \ \pi_{D'}\eta^{D'}, \label{RI}
\end{equation}
which implies that $\pi_{D'}$ and $\eta_{D'}$ are not proportional.

In the case of a spin-frame we have the following normalisation $\pi_{D'}\eta^{D'}=1$, and eq.(\ref{RI}) becomes
\be
R^{\alpha\beta}\mathbb{I}_{\alpha\beta} =  2. \label{normRI}
\ee

Let $Q$ and $R$ be two points in $\mathbb{M}$ given by the position vectors $q^a$ and $r^a$. The distance between these two is obtained as follows
\be \frac{4Q^{\alpha\beta}R_{\alpha\beta}}{Q^{\gamma\delta}\mathbb{I}_{\gamma\delta}R^{\rho\sigma}\mathbb{I}_{\rho\sigma}}=-(q^a-r^a)(q_a-r_a), 
\ee
and if the 2-valence twistors are normalised according to eq.(\ref{normRI}), we have
\be
Q^{\alpha\beta}R_{\alpha\beta}=-(q^a-r^a)(q_a-r_a). \label{horos_twist}
\ee

In terms of the multiparticle space-time algebra, we can identify the infinity twistor with a relativistic massless singlet state. In section \ref{6-d MSTA} we chose $\epsd$ to represent the point at the origin, which leaves us with its complex conjugate $\eps$ as a good candidate for the point at infinity. 

We therefore decide to take as the infinity twistor for Minkowski space-time
\be
I_{\text{M}}=\eps.
\ee
The condition to obtain a finite point, which corresponds to eq.(\ref{RI}), becomes in this formalism
\be
(I_{\text{M}},R^{12})_s=\frac{1}{2}\pietac,
\ee
while in terms of the construction of the 6-d point $\psi_r$ given in section \ref{6-d MSTA}, we have
\be
(I_{\text{M}},\psi_r)_s=\frac{1}{2} (V-W).
\ee
These two equations confirm the relation given by (\ref{normVW}).

Let us now recover the distance function. 
Let $Q^{12}_N$ and $R^{12}_N$ represent the 2-valence twistors encoding the position vectors of the points $q$ and $r$. The distance between these points is given by
\be
-\frac{(Q^{12}_N,R^{12}_N)_s}{2(I_{\text{M}},Q^{12}_N)^*_s(I_{\text{M}},R^{12}_N)_s}=(q-r)\cdot (q-r).
\ee
And if a spin-frame is considered, we can apply a normalisation condition given by $\pietac=1$, which is equivalent to $V-W=1$, to get 
\be
(Q^{12}_N,R^{12}_N)_s=-\half(q-r)\cdot (q-r). \label{QR_mink}
\ee
Note that this relation is identical to its analogue in the single particle space
\be
F_E(q)\cdot F_E(r)=-\half|q-r|^2
\ee
where $F_E(q)$ and $F_E(r)$ are normalised conformal representations of the position vectors $q$ and $r$ in a flat space (see \cite{book}, \cite{proc} and \cite{twistors1}).

\subsection{de Sitter}

De Sitter space (see for example \cite{H&E}) is a space with topology: $R^1\times S^3$. It can be visualised in $\mathbb{P}^5$ as a hyperboloid.
To do this, Penrose and Rindler \cite{PRII} use the coordinates of the 6-d pseudo-Euclidean space $\mathbb{E}^6$, given in section \ref{6-d MSTA}, and restrict them to the hyperplane $T=Q$, where $Q$ is a constant. Instead of that hyperplane, we will use $V=Q$, since $V$ has the same signature as $T$ and for future symmetry with anti-de Sitter space. The condition for the infinity twistor is given by
\be
\begin{split}
\mathbb{I}_{\alpha\beta}R^{\alpha\beta} &= 2\frac{V}{Q} \\
 &= \frac{1}{Q}(1-r_ar^a)\pi_{D'}\eta^{D'} . \label{inf_deSitter}
\end{split}
\ee

The distance function between two points in terms of twistors for this space is of the form 
\be
d(p,r)  =  Q\cosh^{-1}\left(1-\frac{P_{\alpha\beta}R^{\alpha\beta} \mathbb{I}_{\gamma\delta}\mathbb{I}^{\gamma\delta}} {P_{\rho\sigma}\mathbb{I}^{\rho\sigma} \mathbb{I}_{\tau\kappa}R^{\tau\kappa}}\right).
\ee
Imposing a spin-frame, i.e. $\pi_{D'}\eta^{D'}=1$, this expression can be simplified using
\be
\mathbb{I}_{\gamma\delta}\mathbb{I}^{\gamma\delta}=\frac{2}{Q^2},
\ee
and eq.(\ref{inf_deSitter}), to get
\be
d(p,r) = Q \cosh^{-1}\left(1-\frac{2 P_{\alpha\beta}R^{\alpha\beta}}
{(1-p_ap^a)(1-r_ar^a)}\right). \label{conv_d(p,r)_dS}
\ee

De Sitter space has a general negative curvature, which allows us to describe it in terms of a hyperbolic space. In \cite{book}, the point at infinity for such a space in the single particle representation, was taken as the sum of the two null vectors introduced to construct the conformal representation of a space. Therefrom, we choose to define the infinity twistor for de Sitter space in the 2-particle representation, as the sum of the two massless singlet states
\be
I_{\text{dS}}=\eps+\epsd.
\ee
The equation for the condition of a finite point is
\be
(I_{\text{dS}},R_N^{12})_s=\frac{1}{2}(1-|r|^2)\pietac,
\ee
and we can confirm that it is of the form of eq.(\ref{inf_deSitter}).

The derivation for the distance function in the single particle space can be found in \cite{book}. This corresponds to
\be
d(q,r)=\cosh^{-1}\left(1-\frac{F_H(q)\cdot F_H(r)}{(e\cdot F_H(q))((e\cdot F_H(r))}\right)
\ee
where $e$ is the point at infinity in the hyperbolic space, $F_H(q)$ is the conformal representation of the point $q$, and $F_H(r)$ is that of $r$, in the hyperbolic space.

In terms of the 2-particle space this is
\be
d(q,r) = \cosh^{-1}\left(1-\frac{(Q^{12}_N,R^{12}_N)_s(I_{\text{dS}},I_{\text{dS}})_s}{(I_{\text{dS}},Q^{12}_N)^*_s(I_{\text{dS}},R^{12}_N)_s}\right).
\ee
If a spin-frame is introduced it can be further simplified to
\be
d(q,r) =  \cosh^{-1}\left(1+\frac{2(q-r)\cdot (q-r)}
{(1-|r|^2)(1-|q|^2)}\right), \label{d(q,r)_dS}
\ee
since the infinity twistor is such that
\be
(I_{\text{dS}},I_{\text{dS}})_s=1.
\ee
Note that eq.(\ref{d(q,r)_dS}) is in agreement with the definition of a distance function in a hyperbolic space (see \cite{Brannan} and \cite{book}). Furthermore, eq.(\ref{conv_d(p,r)_dS}) takes also that form if eq.(\ref{QR_mink}) is employed.

\subsection{anti-de Sitter}

This space has topology: $S^1\times R^3$. Following the same procedure as above, the coordinates are restricted this time to the hyperplane $W=Q$. The relation that determines the condition to obtain a finite point is given in this space by
\be
\begin{split}
\mathbb{I}_{\alpha\beta}R^{\alpha\beta} &= 2\frac{W}{Q} \\
 &= -\frac{1}{Q}(1+r_ar^a)\pi_{D'}\eta^{D'}, \label{inf_antideSitter}
\end{split}
\ee
and the distance function takes the following form
\be
d(p,r)  =  Q\cos^{-1}\left(1-\frac{P_{\alpha\beta}R^{\alpha\beta} \mathbb{I}_{\gamma\delta}\mathbb{I}^{\gamma\delta}} {P_{\rho\sigma}\mathbb{I}^{\rho\sigma} \mathbb{I}_{\tau\kappa}R^{\tau\kappa}}\right). \ee
Setting a spin-frame, and applying the following simplifications  
\be
\mathbb{I}_{\gamma\delta}\mathbb{I}^{\gamma\delta}=-\frac{2}{Q^2},
\ee
and eq.(\ref{inf_antideSitter}), we get
\be
d(p,r) = Q \cos^{-1}\left(1+\frac{2 P_{\alpha\beta}R^{\alpha\beta}}
{(1+p_ap^a)(1+r_ar^a)}\right). \label{conv_d(p,r)_adS}
\ee

The infinity twistor for this space in the multiparticle space-time algebra framework, can be taken as a linear combination of the massless relativistic singlet states as well. We define this to be
\be
I_{\text{adS}}=\epsd-\eps.
\ee
This choice leads to the following equation for the condition of a finite point
\be
(I_{\text{adS}},R_N^{12})_s=-\frac{1}{2}(1+|r|^2)\pietac,
\ee
which we see corresponds to eq.(\ref{inf_antideSitter}).

The distance function for this space can be constructed through the 2-valence twistors as follows
\be
d(q,r)= \cos^{-1}\left(1-\frac{(Q^{12}_N,R^{12}_N)_s(I_{\text{adS}},I_{\text{adS}})_s}{(I_{\text{adS}},Q^{12}_N)^*_s(I_{\text{adS}},R^{12}_N)_s}\right).
\ee
This can be simplified using normalisation conditions, and we get
\be
d(q,r) = \cos^{-1}\left(1-\frac{2(q-r)\cdot (q-r)}
{(1+|r|^2)(1+|q|^2)}\right), \label{d(q,r)_adS}
\ee
where the infinity twistor is such that
\be
(I_{\text{adS}},I_{\text{adS}})_s=-1.
\ee
If eq.(\ref{QR_mink}) is replaced in eq.(\ref{conv_d(p,r)_adS}), this takes the form of eq.(\ref{d(q,r)_adS}). This confirms that the distance functions are equivalent.

Events in different spaces have therefore been successfully reproduced through 2-valence twistors, in the multiparticle space-time algebra. The spaces considered are of cosmological relevance. Further description of these spaces through twistors is given in the next section.

\subsection{Bang twistor}

For some FRW models there is an initial singularity, the \emph{big bang}. This is not the case for anti-de Sitter and de Sitter spaces.

The energy-momentum tensor of the FRW solutions has the form of a perfect fluid (see for example \cite{H&E}). Expansion (or contraction) of the universe arises from the field equations and assuming conservation of energy. If we take a positive density $\mu$, a non-negative pressure $p$ and a null cosmological constant $\Lambda$, assuming the observed expansion of the universe, an initial singularity in space-time arises from the Raychaudhuri equation. This singularity is also present in models with $\mu+3p>0$ and $\Lambda<0, \ =0$, or $>0$ but very small. 

The metric of the general FRW model can be expressed as (see \cite{PRII})
\be
ds^2=dU^2-[R(U)]^2 d\Omega^2
\ee
where $U$ is the cosmic time and $d\Omega^2$ is the metric of a 3-d space of constant curvature $k$:
\be
d\Omega^2= \frac{dq^2+q^2(d\theta^2+\sin^2\theta d\phi^2)}{(1+\frac{1}{4}kq^2)^2} .
\ee 

The flat case $k=0$, corresponds to Minkowski space-time. The twistor accounting for the singularity in the space is the \emph{bang twistor} $B^{\alpha\beta}$. The structure of this space can be fully described with it and the infinity twistor, although not in a unique way. These relate as follows
\be
\mathbb{I}_{\alpha\beta}B^{\alpha\beta}=0, \qquad 
B_{\alpha\beta}B^{\alpha\beta}=4. \label{bang_inf}
\ee	
The inner product of $B^{\alpha\beta}$ with a general point in the space is given by
\be
\frac{1}{\sqrt{8}}B_{\alpha\beta}R^{\alpha\beta}=T
\ee
describing the expected singularity at the initial time $T=0$.

The above relation is telling us that the twistor is in direct connection with the \emph{time} variable. For this reason, we choose to identify it in the geometric algebra framework with the vector $\gam_0$. In its 2-particle representation it has the following form
\be
B=-\half(\gamma_0^1 \eps I\gamma_3^1+\gamma_0^2 \eps I\gamma_3^2),
\ee
and the singularity at the origin is obtained as expected
\be
(B,R_N^{12})_s=-\half T=-\half t \pietac. 
\ee
We note that the singularity is in the past for an observer at a general position $r$. 

The bang and infinity twistors have the following relations
\be
(I_{\text{M}},B)_s=0 , \quad \text{and} \quad (B,B)_s=\frac{1}{4},
\ee
which are in agreement with eq.(\ref{bang_inf}) (up to a scalar factor).

For the two other cases, $k=1$ and $k=-1$, the structure of the space can be given by two twistors such that
\begin{alignat}{2}
\text{for }k&=1, & \qquad \mathbb{I}_{\alpha\beta}\mathbb{\bar{I}}^{\alpha\beta}&=2, \\
\text{for }k&=-1, & \qquad \mathbb{I}_{\alpha\beta}\mathbb{J}^{\alpha\beta}&=2 . \label{k=-1_1}
\end{alignat}
For $k=1$, $\mathbb{I}^{\alpha\beta}$ and $\mathbb{\bar{I}}^{\alpha\beta}$ are `complex conjugate' simple skew twistors, and for $k=-1$, $\mathbb{I}^{\alpha\beta}$ and $\mathbb{J}^{\alpha\beta}$ are real simple skew twistors, and they can be defined such that
\be
B_{\alpha\beta}=\mathbb{I}_{\alpha\beta}+\mathbb{\bar{I}}_{\alpha\beta} \quad (k=1), \qquad
B_{\alpha\beta}=\mathbb{I}_{\alpha\beta}+\mathbb{J}_{\alpha\beta} \quad (k=-1). \label{k=+-1}
\ee
In terms of coordinates, these new twistors are given by
\begin{alignat}{2}
T-iV&=\frac{1}{\sqrt{2}}\mathbb{I}_{\alpha\beta}R^{\alpha\beta}, \qquad
T+iV&=\frac{1}{\sqrt{2}}\mathbb{\bar{I}}_{\alpha\beta}R^{\alpha\beta} \qquad &(k=1),\\
T-W&=\frac{1}{\sqrt{2}}\mathbb{I}_{\alpha\beta}R^{\alpha\beta}, \qquad
T+W&=\frac{1}{\sqrt{2}}\mathbb{J}_{\alpha\beta}R^{\alpha\beta} \qquad &(k=-1) .
\end{alignat}
In order to find their GA equivalent, let us rewrite these in terms of 4-d components
\begin{alignat}{2} \mathbb{I}_{\alpha\beta}R^{\alpha\beta}&=\sqrt{2}\left\{t-\frac{i}{2}(1-r_ar^a)\right\}\pi_{D'}\eta^{D'}, \qquad &(k=1),\\
\mathbb{I}_{\alpha\beta}R^{\alpha\beta}&=\sqrt{2}\left\{t+\frac{1}{2}(1+r_ar^a)\right\}\pi_{D'}\eta^{D'}, \qquad &(k=-1) . \label{k=-1_2}
\end{alignat}

The case where $k=1$, can be treated as a space of de Sitter type, which although it has general negative curvature, its spatial curvature is positive. In terms of GA the infinity twistor for such a space is given by: $I_{\text{dS}}=\eps+\epsd$. Therefore, the twistor $\mathbb{I}$ can be defined in terms of a combination of that object and the bang twistor. If we choose
\beal
\mathbb{I}&=\frac{1}{2}B+\frac{1}{4}(\eps+\epsd)J,\\
\mathbb{\bar{I}}&=\frac{1}{2}B-\frac{1}{4}(\eps+\epsd)J,
\end{align}
where $J$ corresponds to the unit imaginary in the MSTA (given by eq.(\ref{J})), we obtain
\be
(\mathbb{I},R_N^{12})_s=\frac{1}{4}\left\{-t-\frac{1}{2}(1-|r|^2)I\sigma_3^1\right\}\pietac,
\ee
and
\be
B=\mathbb{I}+\mathbb{\bar{I}},
\ee
which is in agreement with the conventional treatment. Note that this formulation introduces a complexified space with the presence of $J$. Therefore, we get the following relations for the inner products
\be
(\mathbb{\bar{I}},\mathbb{I})_s=0, \quad \text{and} \quad (\mathbb{I},\mathbb{I})_s=\frac{1}{8},
\ee
since $\mathbb{\bar{I}}$ represents the complex conjugate to $\mathbb{I}$.

Finally, the case where $k=-1$ is equivalent to the anti-de Sitter space. This has negative spatial curvature and its infinity twistor is given by $I_{\text{adS}}=\epsd-\eps$. This allows us to make the following choice for the skew twistors
\be
\mathbb{I}=\frac{1}{2}B-\frac{1}{4}(\epsd-\eps),
\ee
and
\be
\mathbb{J}=\frac{1}{2}B+\frac{1}{4}(\epsd-\eps).
\ee
This choice leads to the following relations
\beal
(\mathbb{I},R_N^{12})_s=\frac{1}{4}\left\{-t+\frac{1}{2}(1+|r|^2)\right\}\pietac,\\
(\mathbb{I},\mathbb{J})_s=\frac{1}{8}, \qquad \text{and} \qquad B=\mathbb{I}+\mathbb{J},
\end{align}
which  up to a scalar factor are in agreement with eq.(\ref{k=-1_1}), (\ref{k=+-1}) and (\ref{k=-1_2}).
Furthermore, $\mathbb{I}$ and $\mathbb{J}$ are null twistors.\\

It is interesting to see that these FRW spaces can be fully described by 3 entities from quantum mechanics: $\gamma_0$, the relativistic singlet state $\eps$ and its complex conjugate $\bar{\eps}$. No direct application of these cosmological twistors seems to emerge, but a clear link between different areas suggests that this is a fruitful path worth exploring.

\section{Conclusions}

In paper \cite{twistors1}, we showed that the introduction of the position dependence into the 4-d spinor, in order to define a 1-valence twistor within geometric algebra, has a deeper structure. It arose from the translation of the point at the origin to a general position vector $r$, in the conformal space. In this paper, we verified the consistency of our approach, by deriving \emph{events} in the space-time, through the outer product of 1-valence twistors. Furthermore, we showed that the resulting 2-valence twistor representing such an event, corresponds as well to the point at the origin translated to a general position vector $r$ in the conformal space. In addition, further structure emerged as the result of expressing all the objects in terms of the multiparticle space-time algebra. 
In particular, the conformal elements were represented in terms of relativistic quantum entities, that enabled us to establish a link between a point in the space and a wave-function for a spinless particle. 
We showed that the reality condition that ensures that the 2-valence twistor represents a point that belongs to Minkowski space-time, and therefore represents an \emph{event}, corresponds to the condition within our formalism for the scalar Klein-Gordon field in the wave-function to be \emph{uncharged}. 
If the field has a \emph{charge}, the 2-valence twistor corresponds to the conformal representation of a point that is no longer an event, but a point in the \emph{complexified space-time}.

Following the spirit of the twistor theory, we also constructed more general spaces relevant for cosmology: Minkowski, de Sitter and anti-de Sitter, through the introduction of an \emph{infinity twistor} for each space. 
We found that in our formalism, these 2-valence twistors are given by linear combinations of the massless projections of a relativistic singlet state. Within the 2-particle representation of the conformal geometric algebra, each of these projections correspond to the conformal representation of the point at infinity and the point at the origin in a flat space. The point at infinity in spaces with general positive curvature, or general negative curvature, is as a result constructed as a linear combination of the above mentioned projections. These results have implications for the construction of the distance function. 
We found that this is expressed in terms of the quantum scalar product defined in the multiparticle space-time algebra. 
This product normally takes place between two quantum states, and in this case, the r\^ole of the quantum states is represented by the infinity twistor and the conformal position vector in the 2-particle space.

Further properties of Friedmann-Robertson-Walker spaces were also investigated. For example, for spaces that contain an initial singularity, the big bang, we defined the \emph{bang twistor} within geometric algebra. 
For the ones that do not, the necessary twistors needed to describe the spaces were also given in terms of geometric algebra.  
Note however, that this formalism does not seem to bring any advantages or new applications in cosmology. 
Nevertheless, the close relationship that is found within the multiparticle space-time algebra, between conformal representations of points in real and complexified flat spaces, and relativistic wave-functions, tells us that there is a path worth exploring for the construction of general wave-functions in a unified way.

\section*{Acknowledgements}

Elsa Arcaute would like to acknowledge the financial support of CONACyT.

%\bibliographystyle{plain}
%\bibliography{/home/ea235/biblio,/home/ea235/articles}

\begin{thebibliography}{10}

\bibitem{twistors1}
E.~Arcaute, A.N. Lasenby, and C.J.L. Doran.
\newblock A representation of twistors within geometric ({C}lifford) algebra.
\newblock Submitted to {\em J. Phys. A-Math Gen}, math-ph/0603037, 2006.

\bibitem{Bandos_etal'05}
I.~Bandos, X.~Bekaert, J.A. {de Azc\'arraga}, D.~Sorokin, and M.~Tsulaia.
\newblock Dynamics of higher spin fields and tensorial space.
\newblock {\em J. High Energy Phys.}, 5(31), 2005.
\newblock hep-th/0501113.

\bibitem{Bandos_etal'04}
I.~Bandos, P.~Pasti, D.~Sorokin, and M.~Tonin.
\newblock Superfield theories in tensorial superspaces and the dynamics of
  higher spin fields.
\newblock {\em J. High Energy Phys.}, 11(23), 2004.
\newblock hep-th/0407180.

\bibitem{Barg-Wigner}
V.~Bargmann and E.~Wigner.
\newblock Group theoretical discussion of relativistic wave equations.
\newblock {\em Proc. Nat. Sci. (USA)}, 34:211--223, 1948.

\bibitem{Bette'96}
A.~Bette.
\newblock Directly interacting massless particles - a twistor approach.
\newblock {\em J. Math. Phys.}, 37(4):1724--1734, 1996.

\bibitem{Bette'04}
A.~Bette.
\newblock Twistors, special relativity, conformal symmetry and minimal coupling
  - a review.
\newblock {\em Int. J. Geom. Meth. Mod. Phys.}, 2:265--304, 2005.
\newblock hep-th/0402150.

\bibitem{Bette_etal'04}
A.~Bette, J.A. {de Azc\'arraga}, J.~Lukierski, and C.~Miquel-Espanya.
\newblock Massive relativistic particle model with spin and electric charge
  from two-twistor dynamics.
\newblock {\em Phys. Lett. B}, 595:491--497, 2004.
\newblock hep-th/0405166.

\bibitem{Brannan}
D.A. Brannan, M.F. Espleen, and J.J. Gray.
\newblock {\em Geometry}.
\newblock Cambridge University Press, 1999.

\bibitem{book}
C.J.L Doran and A.N. Lasenby.
\newblock {\em Geometric Algebra for Physicists}.
\newblock Cambridge University Press, 2003.

\bibitem{2}
C.J.L. Doran, A.N. Lasenby, and S.F. Gull.
\newblock States and operators in the spacetime algebra.
\newblock {\em Found. Phys.}, 23(9):1239, 1993.

\bibitem{STA&e}
C.J.L Doran, A.N. Lasenby, S.F. Gull, S.S. Somaroo, and A.D. Challinor.
\newblock Spacetime algebra and electron physics.
\newblock {\em Adv. Imag. \& Elect. Phys.}, 95:271--386, 1996.
\newblock quant-ph/0509178.

\bibitem{H&E}
S.W. Hawking and G.F.R. Ellis.
\newblock {\em The Large Scale Structure of Space-Time}.
\newblock Cambridge University Press, 1973.

\bibitem{proc}
A.N. Lasenby, C.J.L. Doran, and E.~Arcaute.
\newblock Applications of geometric algebra in electromagnetism, quantum theory
  and gravity.
\newblock In R~Ab{\l}amowicz, editor, {\em Clifford Algebras, Applications to
  Mathematics, Physics, and Engineering}, pages 467--489. Birkh{\"a}use, 2004.

\bibitem{V}
A.N. Lasenby and J.~Lasenby.
\newblock Applications of geometric algebra in physics and links with
  engineering.
\newblock In E.~Bayro and G.~Sobczyk, editors, {\em Geometric algebra: {A}
  geometric approach to computer vision, neural and quantum computing, robotics
  and engineering}, pages 430--457. Birkh{\"{a}}user, 2000.

\bibitem{PRII}
R.~Penrose and W.~Rindler.
\newblock {\em Spinors and space-time, Volume II: spinor and twistor methods in
  space-time geometry}.
\newblock Cambridge University Press, 1986.

\end{thebibliography}

\end{document}